\documentclass[useAMS,usenatbib]{mn2e}
\usepackage{times}
\usepackage{graphicx}

\usepackage[pdftex,pdfpagemode={UseOutlines},bookmarks,bookmarksopen,colorlinks,linkcolor={blue},citecolor={blue},urlcolor={red}]{hyperref}

\title[GM\,2-4 region]{GM\,2-4 - a signpost for low and intermediate mass star formation{\thanks{Based on observations collected at the German-Spanish Astronomical Centre, Calar Alto, operated jointly by the Max-Planck-Institut f\"{u}r Astronomie (MPIA) in Heidelberg, Germany, and the Instituto de Astrof\'\i sica de Andaluc\'\i a (CSIC) in Granada/Spain. Includes observations made at the Byurakan Astrophysical Observatory, Byurakan, Armenia.}}}
\author[T. Khanzadyan et al.]{Tigran Khanzadyan$^{1,2,3}$\thanks{e-mail: tkhanzadyan@mpifr-bonn.mpg.de}, 
Tigran A. Movsessian$^{3}$,
Chris J. Davis$^{4,5}$,
\newauthor
Tigran Yu. Magakian$^{3}$,
Roland Gredel$^{6}$,
Elena H. Nikogossian$^{3}$\\
$^{1}$Max-Plank Institut f\"{u}r Radioastronomie, Auf dem H\"{u}gel 69, D-53121 Bonn, Germany\\
$^{2}$Centre for Astronomy, Department of Experimental Physics, National University of Ireland in Galway, Galway, Ireland\\
$^{3}$Byurakan Astrophysical Observatory, 378433 Aragatsotn reg.,Armenia\\
$^{4}$Joint Astronomy Centre, 660 North A`oh\=ok\=u Place, University Park, Hilo, Hawaii 96720, USA\\
$^{5}$Astrophysics Division, NASA HQ, 300 E Street SW, Mail Stop 3Y28 Washington DC 20546, USA\\
$^{6}$Max-Planck Institut f\"{u}r Astronomie, K\"{o}nigstuhl 17, D-69117 Heidelberg, Germany}

\begin{document}

\date{Accepted : Received ...}

\pagerange{\pageref{firstpage}--\pageref{lastpage}} \pubyear{2011}

\maketitle

\label{firstpage}

\begin{abstract} 

We present a multi-wavelength study of the region towards the GM\,2-4 nebula and the nearby source IRAS\,05373+2340. Our near-infrared H$_2$ 1-0 S(1) line observations reveal various shock-excited features which are part of several bipolar outflows. We identify candidates for the driving sources of the outflows from a comparison of the multi-waveband archival data-sets and SED modelling. The SED spectral slope ($\alpha^{IRAC}$) for all the protostars in the field was then compared with the visual extinction map. This comparison suggests that star formation is progressing from NE to SW across this region.
\end{abstract}


\section{Introduction} 
\label{sec:intro}

Since 1998, the Byurakan observatory in Armenia has been engaged in surveys of nebular objects in dark molecular clouds and star formation regions. With its 2.6m telescope, imaging in narrow band filters and simultaneous searches of H$\alpha$ emission stars are being performed. New HH outflows and star forming regions have been found \citep[see][and references therein]{2008Ap.....51..181M}. An exploration of the region with GM\,1-64 (RNO\,53)  \citep{1977SvAL....3...58G,1980AJ.....85...29C} and the GM\,2-4 nebulae \citep{1977DoArm..64..104G} at optical wavelengths has lead to the discovery of several HH objects as well as numerous H$\alpha$ emission stars \citep{2009Ap.....52..501N}. 

\begin{figure*}
	\centering
	\includegraphics[width=15cm]{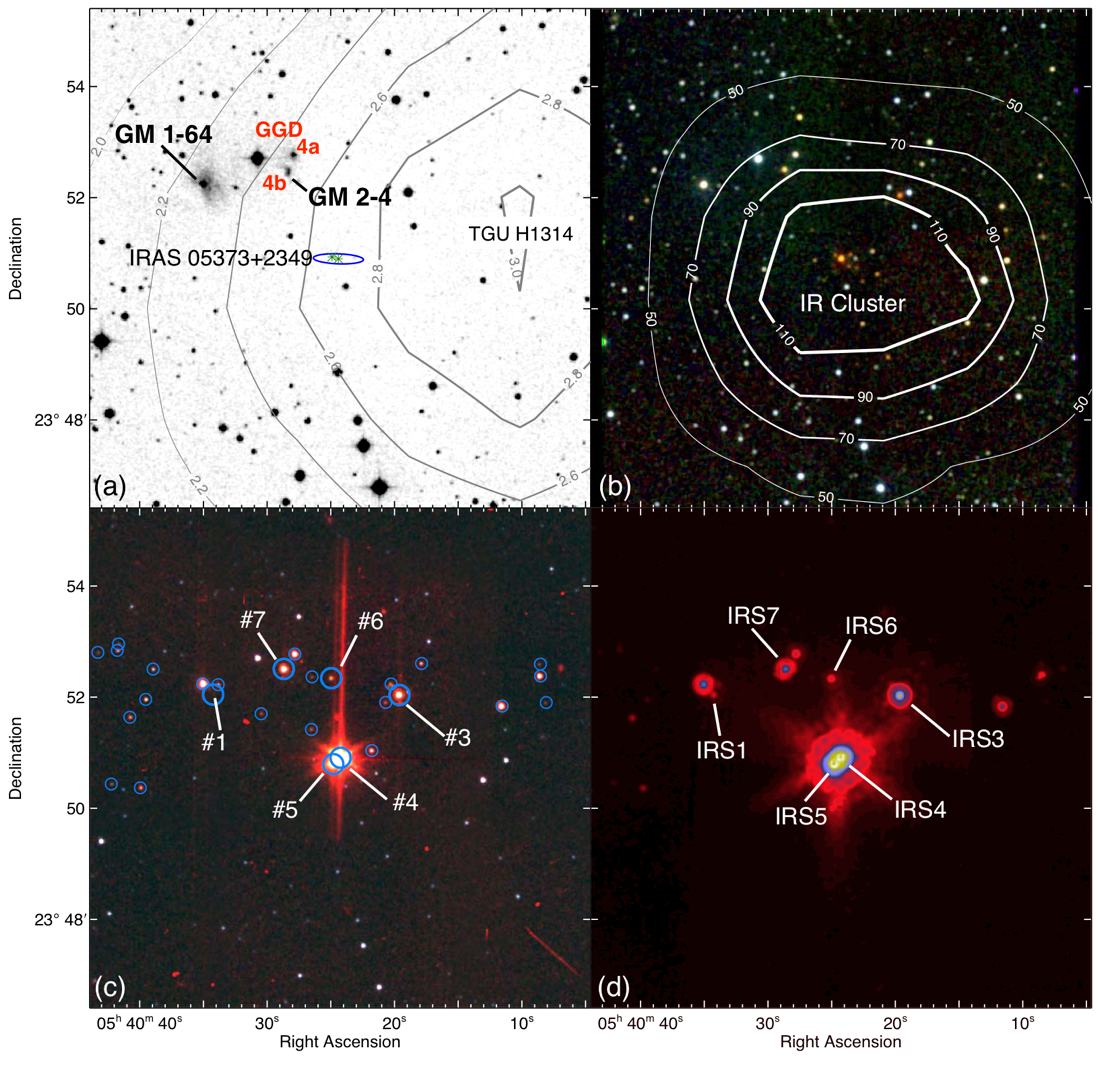}
	\caption{General field including GM\,1-64 and GM\,2-4 centred on IRAS\,05373+2349. \textbf{Panel (a)} is DSS2 red image of the region overlaid with a visual extinction contours obtained from \citet{2005PASJ...57S...1D} peaking at TGU\,H1314. The solid-line ellipse indicates the positional uncertainty of IRAS\,05373+2349 and the green asterisks mark the positions of masers. \textbf{Panel (b)} is the 2MASS false colour composite view of the region constructed from J (blue), H (green) and K$_s$ (red) bands. Overlaid are the \emph{IRAS} 100$\mu$m contours at 50, 70, 90 and 110 MJy/sterad. \textbf{Panel (c)} is \emph{Spitzer} IRAC colour composite image constructed from 3.6$\mu$m (blue), 4.5$\mu$m (green) and 8.0$\mu$m (red) bands. Stars marked with big and small circles are sources reported by \citet{2009ApJS..184...18G} and classified as Class\,I protostars and Class\,II pre-main-sequence stars respectively. \textbf{Panel (d)} is \emph{Spitzer} MIPS 24$\mu$m band view of the region where the same Class\,I sources are marked with the nomenclature used in our study (see Sect\,\ref{irs-nomencl}).}
	\label{fig:genfield}
\end{figure*}

GM\,2-4 and the nearby IRAS\,05373+2349 are located in the centre of the [KOY98]183.7-03.6 molecular cloud which has been detected in $^{13}$CO by \citet{1998ApJS..117..387K}. The authors provide a distance estimate of $\sim$2kpc. \citet{2005PASJ...57S...1D} identified a peak of optical extinction (TGU\,H1314 in Fig.\,\ref{fig:genfield}) from DSS1 plates which is located about 1.5-2.0\arcmin westwards of IRAS\,05373+2349 (cf. Fig.\,\ref{fig:genfield}a). This peak appears to be associated with the Shajn\,147 supernova remnant \citep{1986A&A...169..281C}. The distance to the SNR progenitor PSR\,J0538+2817 is obtained from a precise parallax measurement and is  given as 1.3kpc \citep{2009ApJ...698..250C}. These distances are smaller than the early estimates of 1.7 kpc based on the kinetic velocities of radio lines \citep[eg.][]{1980ApJ...235..845R}, but they agree with subsequent radio observations presented by \citet{1996A&A...308..573M} who infer a distance of 1.17 kpc from an updated Galactic rotational model \citep{1993A&A...275...67B}

GM\,2-4 was discovered on DSS1 plates \citep{1977DoArm..64..104G} and is classified as a Herbig-Haro-Like object (HHL\,26) by \citet{1978ApJ...224L.137G} and later renamed GGD\,4 by \citet{1987RMxAA..15...53G}. Subsequent near-infrared observations \citep{1988MNRAS.232..497C} were able to separate this object into two separate red nebulous objects namely GGD\,4a and GGD\,4b coinciding with the position of GM\,2-4 (Fig.\,\ref{fig:genfield}a). A recent continuum and narrow-band (H$\alpha$ and [SII]) optical study by \citet{2009Ap.....52..501N} concludes that these objects (GM\,2-4 and GGD\,4a) are reflection nebulae and do not exhibit any Herbig-Haro characteristics. The near-infrared H$_2$ 1-0 S(1) line survey of \citet{1983ApJ...265..864S} did not find  molecular hydrogen line emission, but this is likely due to the limited sensitivity of the infrared bolometer used in that study. The detection of an H$_2$O maser by \citet{1980ApJ...235..845R} and the detection of a cold IRAS point source 05373+2349\footnote{IRAS catalogue of Point Sources, Version 2.0 (IPAC 1986)} about 1.5\arcmin south-west from the centre of GM\,2-4 (cf. Fig.\ref{fig:genfield}a) indicated that active star-formation occurs in the region. 

\citet{1986A&A...169..281C} detected high-velocity blue and red-shifted wings in the $^{12}$CO emission which pointed to a bi-polar outflow driven by IRAS\,05373+2349. \citet{1988MNRAS.232..497C} obtained photometric measurements in near- to mid-infrared bands of a red object coincident with the position of the IRAS\,05373+2349 using an infrared bolometer. The same red object was resolved into an infrared stellar cluster by \citet{1994ApJS...94..615H} (cf. Fig.\ref{fig:genfield}b). This infrared cluster is also associated with 450$\mu$m continuum emission \citep{1995MNRAS.276.1024J} and with an ammonia core \citep{1996A&A...308..573M}. More recently, 3.4cm and 3.6cm continuum emission and HCO$^+$ (1-0), H$^{13}$CO$^+$ (1-0) and H$^{13}$CN (1-0) line emission were detected towards the cluster \citep{2002ApJ...570..758M}. A fit to the spectral energy distribution (SED) of  IRAS\,05373+2349 from 8 - 1200 $\mu$m by \citet{2008A&A...481..345M} was interpreted in terms of a zero-age main-sequence B5 star of 26 M$_{\sun}$ for the entire cluster of stars. The study by \citet{2009ApJS..184...18G} using \emph{Spitzer} addressed the evolutionary stages of individual members in the IR cluster, and found that its members were a mixture of Class\,I protostars and Class\,II pre-main-sequence stars with optically thick disks (cf. Fig.\ref{fig:genfield}c). \citet{2010MNRAS.404..661V} reported some very faint near-infrared H$_2$ 1-0 S(1) line knots around the position of IRAS\,05373+2349. 

These studies demonstrate that the area around GM\,2-4 is an active and on-going star formation region. We have thus decided to combine previously published data on the region with our wide-field near-infrared survey of the region in the outflow tracing H$_2$ 1-0 S(1) line in order to reveal the extent of outflow activity, identify their driving sources and examine the star formation processes of the studied field. In Section\,\ref{sec:obs} we present the observations and the data analysis. In Section\,\ref{sec:resdisc} we present the results from our observations, discuss the possible outflows and present SED modelling on suspected driving sources. Finally, in Section\,\ref{sec:sum} we summarise our findings and present some future prospectives on the project.

\begin{figure*}
	\centering
		\includegraphics[width=17.5cm]{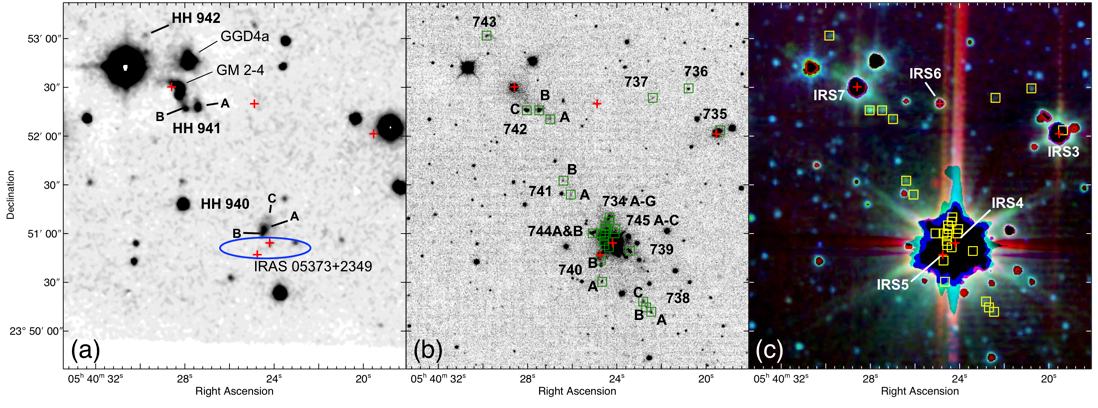}
	\caption{Close-up view of GM\,2-4 and IRAS\,05373+2349 including the region where new molecular hydrogen emission line objects (MHOs) are found. Class\,I sources reported by \citet{2009ApJS..184...18G} are marked with a crosses on all panels. \textbf{Panel (a)} is optical [SII] image where Herbig-Haro objects discovered by \citep{2009Ap.....52..501N} are marked along with the GM\,2-4, GGD\,4a and IRAS\,05373+2349 error ellipse. \textbf{Panel (b)} is near-infrared H$_2$ 1-0 S(1) line+continuum image where the MHOs are marked with squares. \textbf{Panel (c)} shows \emph{Spitzer} IRAC colour-composite image constructed from 3.6$\mu$m (blue), 4.5$\mu$m (green) and 8.0$\mu$m (red) bands with positions of newly detected MHOs marked with squares for a comparison. For this particular case the brightest (mostly saturated) stars in the field were truncated in order to stretch the colour scheme and reveal relatively fainter objects.}
	\label{fig:mainfigure}
\end{figure*}

\section{Observations and Data Analysis} 
\label{sec:obs}

The following discussion makes frequent use of the optical observations by \citet{2009Ap.....52..501N} - see that paper for a comprehensive discussion of the optical dataset.

\subsection{Near-infrared Data}
\label{obs_nir}

Observations in the near-Infrared were carried out during the night of October 11, 2004 using the Omega 2000 prime focus camera \citep{2003SPIE.4841..343B} at the Calar Alto 3.5m telescope. Omega 2000 is equipped with a 2048 $\times$ 2048 pixel HAWAII-2 array detector. It provides a pixel scale of 0.45 arcsec/pixel and a total field of view of 15.4 $\times$ 15.4 arcmin$^2$ on the sky. Images were obtained in the H$_2$ 1-0 S(1) line using a 1\% narrow band filter centred at $\lambda$ = 2.122$\mu$m, and in a 1\% narrow band continuum filter centred at $\lambda$ = 2.143$\mu$m. The total exposure time was 1500s per filter with the average seeing of $\sim$ 1.0\arcsec. The data reduction proceeded through standard routines using CCDPACK and KAPPA data reduction packages developed under the Starlink Project: i.e. the images were dark subtracted, flat fielded, sky subtracted and then combined together to form a mosaic picture of the complete field. In the next step images in $\lambda$ = 2.122$\mu$m and 2.143$\mu$m wavelengths were astrometrically calibrated and subtracted in order to reveal the pure H$_2$ 1-0 S(1) line emission objects in the field.

\subsection{Mid-infrared Data}

We acquired IRAC \citep{2004ApJS..154...10F} and MIPS \citep{2004ApJS..154...25R} data from the \emph{Spitzer} Science Archive covering the area around GM\,2-4. The basic flux calibrated images of the \emph{Spitzer} Science Center (SSC) pipeline were used. Cosmetic corrections, astrometric refinement and the final mosaics were constructed using the MOPEX software \citep{2005ASPC..347...81M}. Photometric measurements were done using the aperture photometry package PHOTOM in GAIA\footnote{GAIA is a derivative of the Skycat catalogue and image display tool, developed as part of the VLT project at ESO. Skycat and GAIA are free software under the terms of the GNU copyright}. Instrumental values (counts/sec) where calibrated using the guidelines presented in SSC home page (i.e. appropriate aperture and pixel-scale correction) and zero-point fluxes by \citet{2005PASP..117..978R} based on Vega-standard magnitudes for 1 DN/s. Our results are in excellent agreement with the values presented in \citet{2009ApJS..184...18G} differing only by a few percent.

\section{Results and Discussion} 
\label{sec:resdisc}

\begin{figure}
	\centering
		\includegraphics[width=8cm]{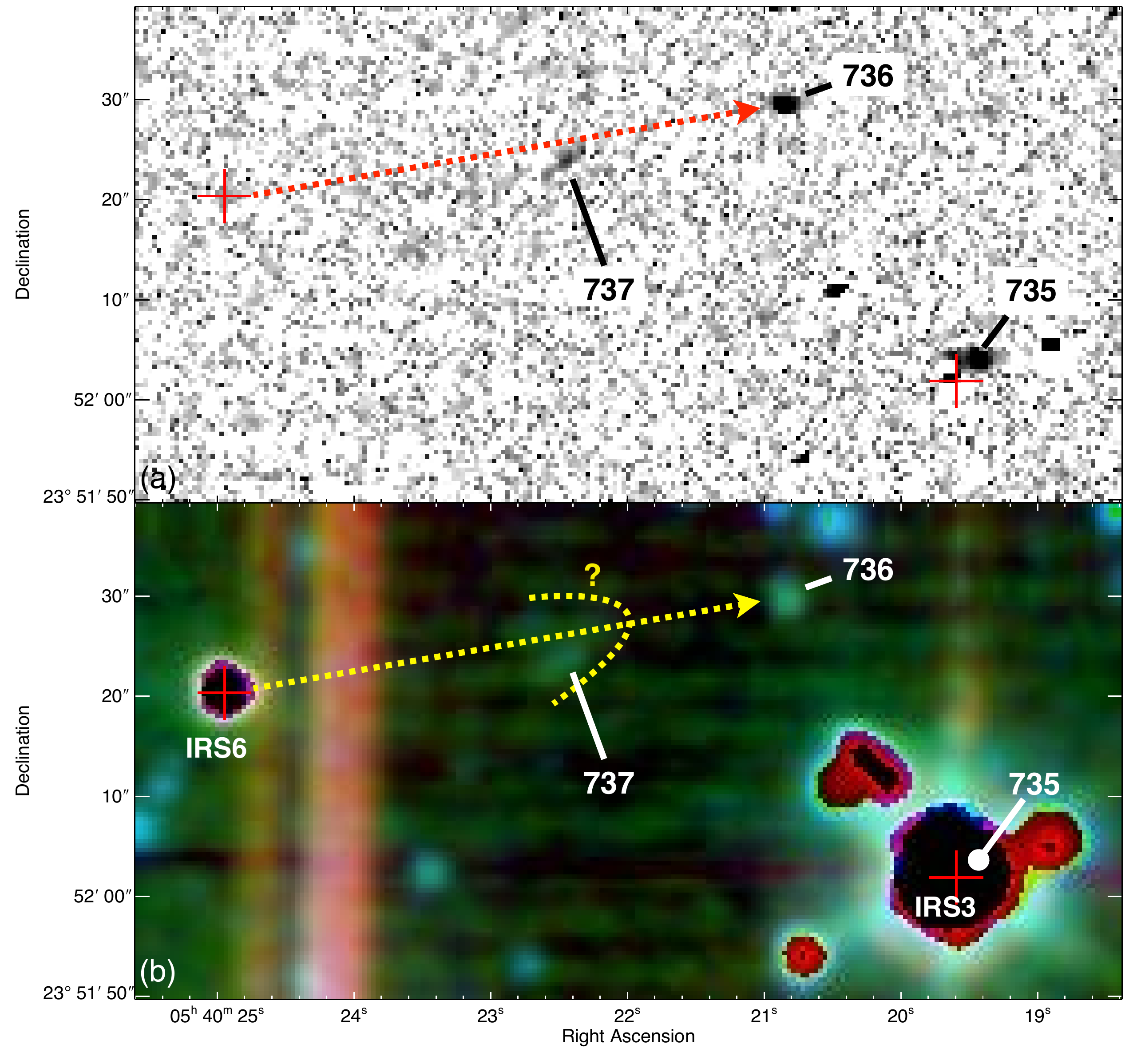}
	\caption{Area containing the MHO\,735, 736 and 737. Positions of IRS objects are marked with the crosses on all panels. Dotted arrow on both panels show the possible outflow direction. \textbf{Panel (a)} shows the H$_2$ 1-0 S(1) emission. \textbf{Panel (b)} is a colour composite image constructed from 3.6$\mu$m (blue), 4.5$\mu$m (green) and 8.0$\mu$m (red) \emph{Spitzer} IRAC bands where the position of MHO\,735 is marked with a filled circle for an easy comparison and a dotted bow-shock outlines the MHO\,737.}
	\label{fig:irs3and6}
\end{figure}

\subsection{Definitions and Nomenclature}
\label{irs-nomencl}

Here we give a brief reasoning on the used nomenclature for the protostellar objects referred in the following sections. We follow the \citet{2009ApJS..184...18G} source numbering with a slight difference aiming at clarifying the overall situation in the immediate vicinity of the GM\,2-4 and IRAS\,05373+2349 objects. \citet{2009ApJS..184...18G} call these objects the \textbf{GGD\,4 cluster} of sources, but this nomenclature in our view is misleading in two ways. Firstly the actual position of the GGD\,4 object reported originally by  \citet{1978ApJ...224L.137G} is about 1.5 arc-minutes to the north-east of the near-infrared cluster discovered by \citet{1994ApJS...94..615H} (Fig.\,\ref{fig:genfield}(b)). Secondly the GGD\,4 object, which were later renamed GGD\,4a and GGD\,4b (GM\,2-4) by \citet{1988MNRAS.232..497C}, are a pair of reflection nebulae. These reflection nebulae are not associated with HH objects; nor are they obviously connected with the maser sources reported by \citet{1980ApJ...235..845R}. We believe that the cluster should be called the  \textbf{IRAS\,05373+2349 cluster} and the numbering of the infrared sources reported by \citet{2009ApJS..184...18G} should be preserved. Therefore in the interest of brevity, and because of the absence of any other clusters in the region we will use the \textbf{IR cluster} and the \textbf{IRS\#} (Infra-Red-Source) nomenclature throughout this work.

\subsection{New Molecular Hydrogen emission line Objects (MHOs)}

Figure\,\ref{fig:mainfigure} presents a part of our region where we detected numerous H$_2$ 1-0 S(1) line emission objects. In order to identify pure H$_2$ 1-0 S(1) line emission objects we subtracted the 2.143$\mu$m narrow-band continuum image from the 2.121$\mu$m line image. For the comparison we include optical [SII] line data \citep{2009Ap.....52..501N} along with the colour-composite figure constructed from \emph{Spitzer} IRAC bands. We detected number of new Molecular Hydrogen emission-line Objects \citep{2010A&A...511A..24D} (MHOs\footnote{\url{http://www.jach.hawaii.edu/UKIRT/MHCat/}} hereafter). Table\,\ref{tab:mho} provides the coordinates, flux in units of 10$^{-18}$~W~m$^{-2}$ arcsec$^{-2}$  and some additional notes related to the newly discovered MHOs. In the following subsections we briefly describe each outflow.

\subsubsection{MHO\,735, 736 and 737}

About 2 arc-minutes west from GM\,2-4 (cf. Fig.\,\ref{fig:mainfigure}a,b) three emission line objects MHO\,735, 736 and 737 were detected. The detailed view of the area in near-infrared H$_2$ 1-0 S(1) emission and the composite constructed from the \emph{Spitzer} IRAC bands including these objects are presented on Fig.\,\ref{fig:irs3and6}a and b.

MHO\,735 is situated very close to the continuum source IRS3 which appears prominently at mid-infrared wavelengths and is most likely the driving source of MHO\,735 based on the positional closeness ($\sim$ 0.018pc on 1.17kpc). We were unable to identify MHO\,735 in the 3.6$\mu$m, 4.5$\mu$m and 8.0$\mu$m \emph{Spitzer} IRAC bands, due to the large point spread function of IRS3 (cf. Fig.\ref{fig:irs3and6}b). In contrast MHO\,736 and 737 are detected at 4.5$\mu$m as expected  if the emission originates from shock excited H$_2$ lines \citep{2005MNRAS.357.1370S,2009ApJ...695L.120Y}. A confirmation of the shocked nature in 4.5$\mu$m would require a spectroscopic follow-up.

A close examination of the colour-composite IRAC view (Fig.\ref{fig:irs3and6}b) reveals the bow-shock shape of the MHO\,737 which suggests the IRS6 as a possible driving source of a flow with a length of at least 0.195pc (Tab.\,\ref{tab:outflows}). The direction of the flow (if confirmed) would also be suitable for inclusion of MHO\,736 as a forward shock making it over 0.325pc long.


\begin{table}
\caption{List of MHOs detected in H$_2$ 1-0 S(1) with cross-identifications with HH objects from \citet{2009Ap.....52..501N} and the knots identified by \citet{2010MNRAS.404..661V} are marked by [VDR]8-\# nomenclature.}
\label{tab:mho}
 \begin{tabular}{lcccl}
        \hline
        \textbf{MHO}& 
        \textbf{R.A.(2000)}& 
        \textbf{Dec.(2000)}&
        \textbf{Flux$^{\dagger}$}&
        \textbf{Notes}\\
        \hline
		734A & 05:40:24.37 & +23:51:10.6 & 25.5 & HH\,940C\\
		734B & 05:40:24.40 & +23:51:07.9 & 23.9 & \\
		734C & 05:40:24.57 & +23:51:05.6 & 12.6 & within [VDR]8-3\\
		734D & 05:40:24.53 & +23:51:03.3 & 27.8 & HH\,940A, [VDR]8-4\\
		734E & 05:40:24.60 & +23:50:59.7 & 43.6 & HH\,940B, [VDR]8-2\\
		734F & 05:40:24.62 & +23:50:56.1 & 7.7  & \\
		734G & 05:40:24.59 & +23:50:53.4 & 11.5 & \\
		735  & 05:40:19.43 & +23:52:04.4 & 13.9 & \\
		736  & 05:40:20.86 & +23:52:30.1 & 11.8 & \\
		737  & 05:40:22.47 & +23:52:24.2 & 4.1  & \\
		738A & 05:40:22.47 & +23:50:12.5 & 8.8  & \\
		738B & 05:40:22.70 & +23:50:15.2 & 4.7  & \\
		738C & 05:40:22.84 & +23:50:18.8 & 9.6  & \\
		739  & 05:40:23.44 & +23:50:49.9 & 18.9 & [VDR]8-6\\
		740A & 05:40:24.68 & +23:50:30.9 & 7.4  & \\
		740B & 05:40:24.75 & +23:50:43.9 & 5.3  & \\
		741A & 05:40:26.12 & +23:51:24.4 & 3.4  & \\
		741B & 05:40:26.45 & +23:51:32.9 & 1.8  & \\
		742A & 05:40:27.06 & +23:52:10.8 & 10.8 & \\
		742B & 05:40:27.55 & +23:52:16.2 & 32.4 & HH\,941A\\
		742C & 05:40:28.11 & +23:52:16.2 & 50.4 & HH\,941B\\
		743  & 05:40:29.94 & +23:53:02.1 & 10.5 & HH\,942\\
		744A & 05:40:25.12 & +23:51:00.6 & 32.2 & [VDR]8-2\\
		744B & 05:40:24.76 & +23:51:00.6 & 12.3 & [VDR]8-3\\
		745A & 05:40:24.10 & +23:51:03.4 & 4.9  & \\
		745B & 05:40:24.14 & +23:51:00.6 & 13.2 & [VDR]8-5\\
		745C & 05:40:24.39 & +23:50:52.1 & 23.2 & [VDR]8-1\\
		\hline
	\end{tabular}
	\par $^\dagger$ Flux in 10$^{-18}$~W~m$^{-2}$ units and the background 1$\sigma$ noise estimate is $1.15 \times 10 ^{-19}$~W~m$^{-2}$ calculated in 6\arcsec\ circular aperture
\end{table}

\subsubsection{MHO\,734, 739, 740, 744 and 745}

\begin{figure*}
	\centering
		\includegraphics[width=13cm]{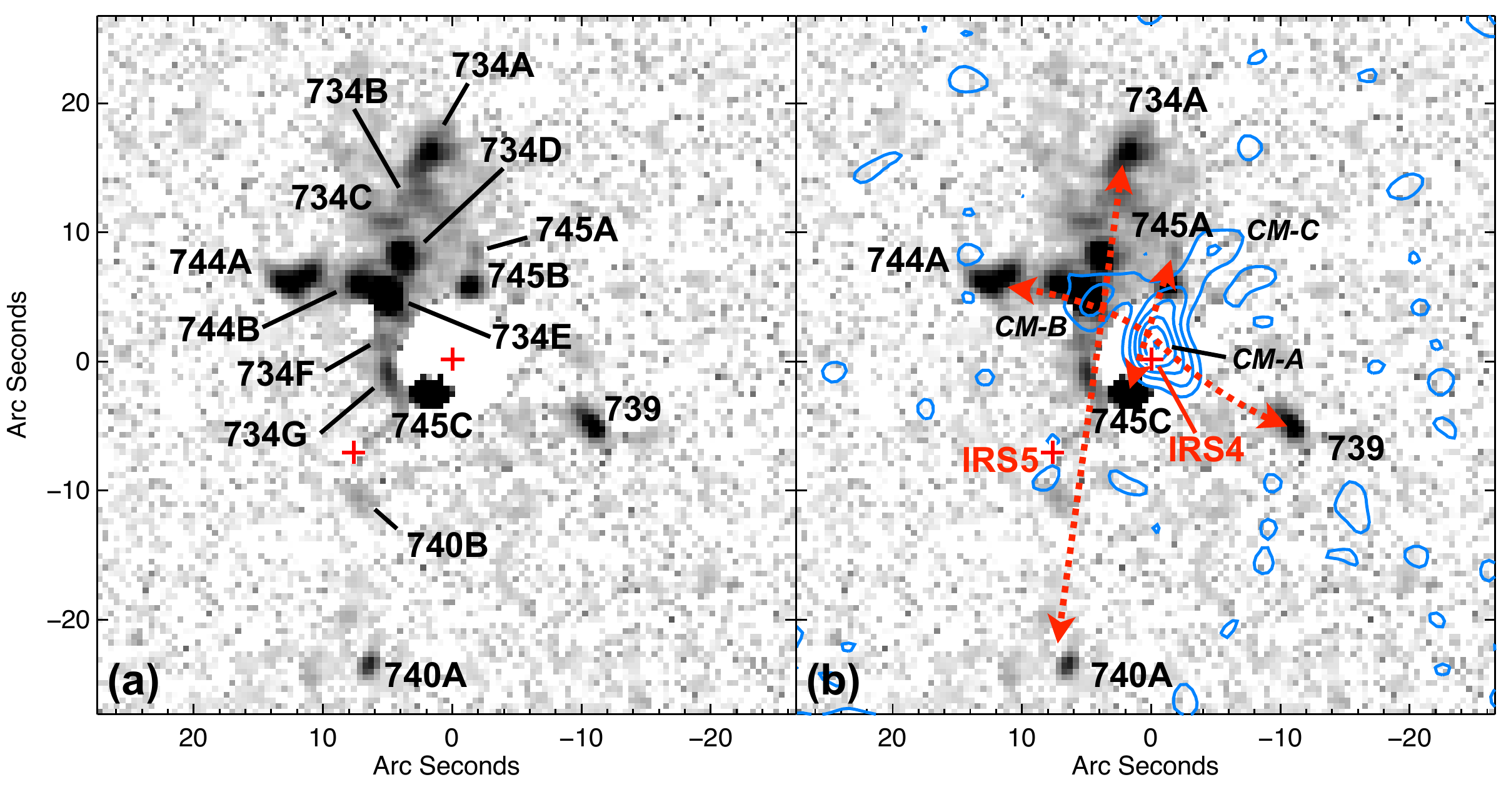}
	\caption{Close-up view in H$_2$ 1-0 S(1) emission of the area surrounding the IRAS\,05373+2349 and the near-infrared cluster where numerous MHOs where identified. Positions of IRS4 and 5 are marked with crosses on all panels. \textbf{Panel (a)} reveals details of MHO\,734, 739, 740, 744 and 745. \textbf{Panel (b)} shows the suggested outflow structure with dotted arrows. Overlaid contours are from the 3.6cm emission starting from 3$\sigma$ and climbing with a 3$\sigma$ step-size and are from the data published by \citet{2002ApJ...570..758M}. Positions of 3.6cm peaks are marked with CM-A, -B and -C.}
	\label{fig:irs4and5}
\end{figure*}

A detailed view of the region around IRAS\,05373+2349 (IRS4 and 5) shows a complex structure of knots and bows in the immediate vicinity of IRS4 (cf. Fig.\,\ref{fig:irs4and5}a). This region has been studied by \citet{2010MNRAS.404..661V} who also detected several faint H$_2$ 1-0 S(1) line features (object 8 in their paper) which coincide with some of the individual knots detected in our study (cf. Tab.\,\ref{tab:mho} for cross-identification). In our deeper images we were able to detect more features and therefore suggest several possible outflows, which are marked on Fig.\,\ref{fig:irs4and5}b and listed in Table\,\ref{tab:outflows}.

An elongated object MHO\,739 located about 10\arcsec\ south-west from IRS4 was also detected by \citet{2010MNRAS.404..661V} who label this feature as 8-6. This object could be connected with the MHO\,744B compact knot (8-3) and MHO\,744A bow-shock (8-2) to form a 0.144pc long bipolar outflow (739-744 outflow) possibly driven by IRS4/CM-A (Fig.\,\ref{fig:irs4and5}b). Next bipolar flow (745flow $\sim$ 0.068pc), perpendicular to the 739-744 outflow, connecting MHO\,745A and B on one side and MHO\,745C on the other side is most likely driven from the very same IRS4/CM-A source \citep{2002ApJ...570..758M}.

The chain of bow-shocks, compact knots and elongated features denoted as MHO\,734A to G (cf. Fig.\,\ref{fig:irs4and5}a) connect with the MHO\,740B and 740A knots forming the third bipolar outflow (734-740 $\sim$ 0.226pc; Tab.\,\ref{tab:outflows}) in this active region (Fig.\,\ref{fig:irs4and5}b). Determination of the true driving source of this bipolar outflow is unrealistic due to the limiting dataset, but one could suggest some candidates based on a simple geometrical placement of the probable candidates. In this respect CM-B \citep{2002ApJ...570..758M} is the most probable driving source for this outflow but one could also consider IRS4/CM-A source because of its apparent closeness and prominence. Lastly the overall outflow direction described by 734-740 (Tab\,\ref{tab:outflows}) coincides with the  $^{12}$CO outflow reported previously by \citet{1986A&A...169..281C}.

\subsubsection{MHO\,738 and 741}

\begin{figure}
	\centering
		\includegraphics[width=8cm]{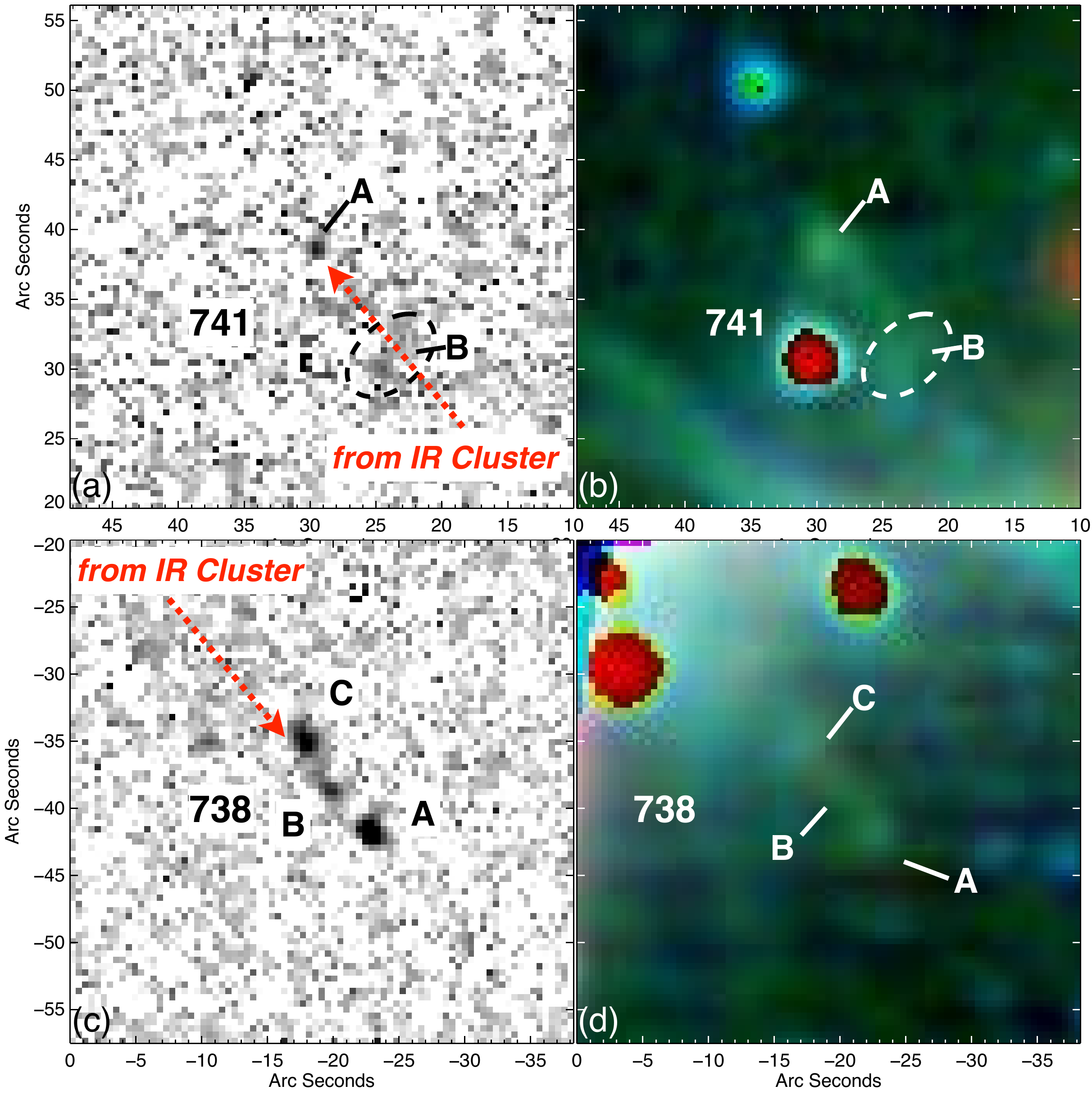}
	\caption{Scaled-up views of MHO\,738 and 741 where the offsets are calculated from IRS4. Dotted-line arrows show the direction of the suggested bipolar outflow. \textbf{Panel (a)} shows the area including MHO\,741 in H$_2$ 1-0 S(1) emission and the \textbf{panel (b)} is the same area in \emph{Spitzer} IRAC composite view constructed from 3.6$\mu$m (blue), 4.5$\mu$m (green) and 8.0$\mu$m (red) bands. \textbf{Panel (c)} and \textbf{panel (d)} show the same wavelengths as in panels (a) and (b) but for the area including the MHO\,738A, B and C.}
	\label{fig:mho738a41}
\end{figure}

Figure\,\ref{fig:mho738a41} gives the details of MHO\,738 and 741 in H$_2$ 1-0 S(1) emission and in \emph{Spitzer} IRAC composite view where the offsets are calculated from IRS4 (cf. Fig.\,\ref{fig:mainfigure}).

About 40\arcsec\ south-west from the IRS4 several bow-shaped features labelled MHO\,738A to C are located (Fig.\,\ref{fig:mho738a41}c and d). The morphology and their placement suggests that one of the sources in the \textbf{IR cluster} drives an outflow creating the MHO\,738\,A-C bow shocks. To the north-east of the \textbf{IR cluster}, again at a distance of about 40\arcsec\, the two faint knots MHO\,741A and 741B are located (Fig.\,\ref{fig:mho738a41}a and b).

MHO\,738 and 741 quite likely are part of yet another bipolar outflow ($\sim$ 0.498pc; Tab.\,\ref{tab:outflows}) driven from the sources in the \textbf{IR cluster}. As a candidate driving source for the 738-741 outflow both IRS4/CM-A and CM-B can be considered.

\subsubsection{MHO\,742 and 743}

\begin{figure}
	\centering
		\includegraphics[width=8cm]{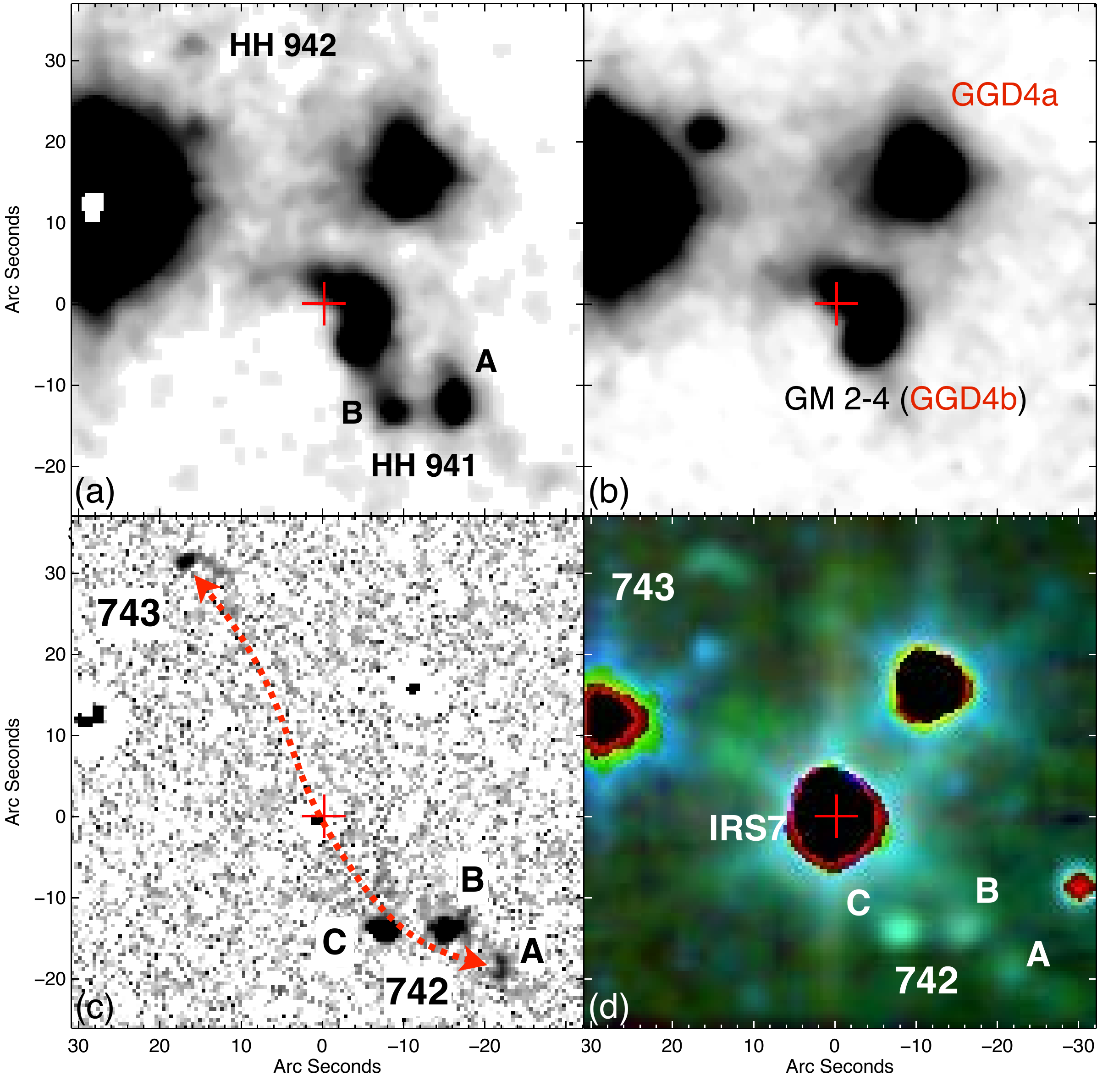}
	\caption{Region around IRS7 including MHO\,742 and 743. \textbf{Panel (a)} is optical [SII] view where HH\,941 and 942 are marked. \textbf{Panel (b)} is optical I-band continuum view where the reflection nebulae GM\,2-4 (GGD\,4b) and GGD\,4a are indicated. \textbf{Panel (c)} is the near-infrared H$_2$ 1-0 S(1) line image where MHO\,742 and 743 are indicated. Dotted arrow shows the suggested bipolar outflow. \textbf{Panel (d)} is \emph{Spitzer} IRAC composite view constructed from 3.6$\mu$m (blue), 4.5$\mu$m (green) and 8.0$\mu$m (red) bands where along with MHO\,742 and 743 IRS7 is also indicated.}
	\label{fig:irs7}
\end{figure}

The area around GM\,2-4 has been studied recently by \citet{2009Ap.....52..501N} who found two Herbig-Haro objects, HH\,941 and HH\,942. Our follow-up on the same region reveal near- to mid-infrared counterparts of these HH objects. Fig.\,\ref{fig:irs7} gives the optical [SII] and I-band, near-infrared H$_2$ 1-0 S(1) and \emph{Spitzer} IRAC views of the region around the optically undetectable IRS7 which illuminates the GM\,2-4 (GGD\,4b) nebula. For completeness we note that GGD\,4a is also illuminated by a near- to mid-infrared source which was classified as a Class\,II pre-main-sequence star by \citet{2009ApJS..184...18G} (cf. Fig\,\ref{fig:genfield}).

About 20\arcsec\ south-west of IRS7 (Fig.\,\ref{fig:irs7}c) the MHO\,742 object composed of several prominent knots (A, B and C) was identified coinciding with the optical HH\,941A and B (cf. Tab.\,\ref{tab:mho}). While knots MHO\,742B and C are less defined the MHO\,742A knot clearly resembles a bow-shock like structure pointing back to IRS7. Similarly about 30\arcsec\ north-east of IRS7 (Fig.\,\ref{fig:irs7}c) lays MHO\,743 which has a clear bow-like morphology pointing back to the IRS7 as well. This makes 742-743 a 0.372pc long bipolar outflow (Tab\,\ref{tab:outflows}) emanating from IRS7.

In addition all features detected in near-infrared H$_2$ 1-0 S(1) line are also clearly detectable in green on the \emph{Spitzer} IRAC composite view (Fig.\,\ref{fig:irs7}d) which suggests that the radiation must be coming from the shock excited H$_2$ lines found in IRAC 4.5$\mu$m band.

\begin{table}
	\caption{Outflows and their driving sources for the discussed region.}
	\label{tab:outflows}
	\begin{tabular}{lcccl}
		\hline
		\textbf{MHO outflow} & 
		\textbf{Source} & 
		\textbf{Size (pc)}$^{\dagger}$ & 
		\textbf{Flux}$^{\star}$ & 
		\textbf{Notes}\\
		\hline
		735flow  & IRS3 & 0.018 & 13.9 & monopolar\\
		736flow  & IRS6 & 0.325 & 15.9 & monopolar\\
		738-741  & IRS4/CM-A & 0.498 & 28.3 & bipolar\\
		739-744  & IRS4/CM-A & 0.144 & 63.1 & bipolar\\
		745flow  & IRS4/CM-A & 0.068 & 41.3 & bipolar\\
		734-740  & CM-B & 0.226 & 165.3 & bipolar\\
		742-743  & IRS7 & 0.372 & 104.1 & bipolar\\
		\hline
	\end{tabular}
	\par $^{\dagger}$ Assuming 1.17kpc distance\\
	$^{\star}$ Combined flux in 10$^{-18}$~W~m$^{-2}$ units
\end{table}

\subsection{SED modelling of IRS Objects}
\label{sec:sedmodels}

\begin{figure*}
	\centering
		\includegraphics[width=17cm]{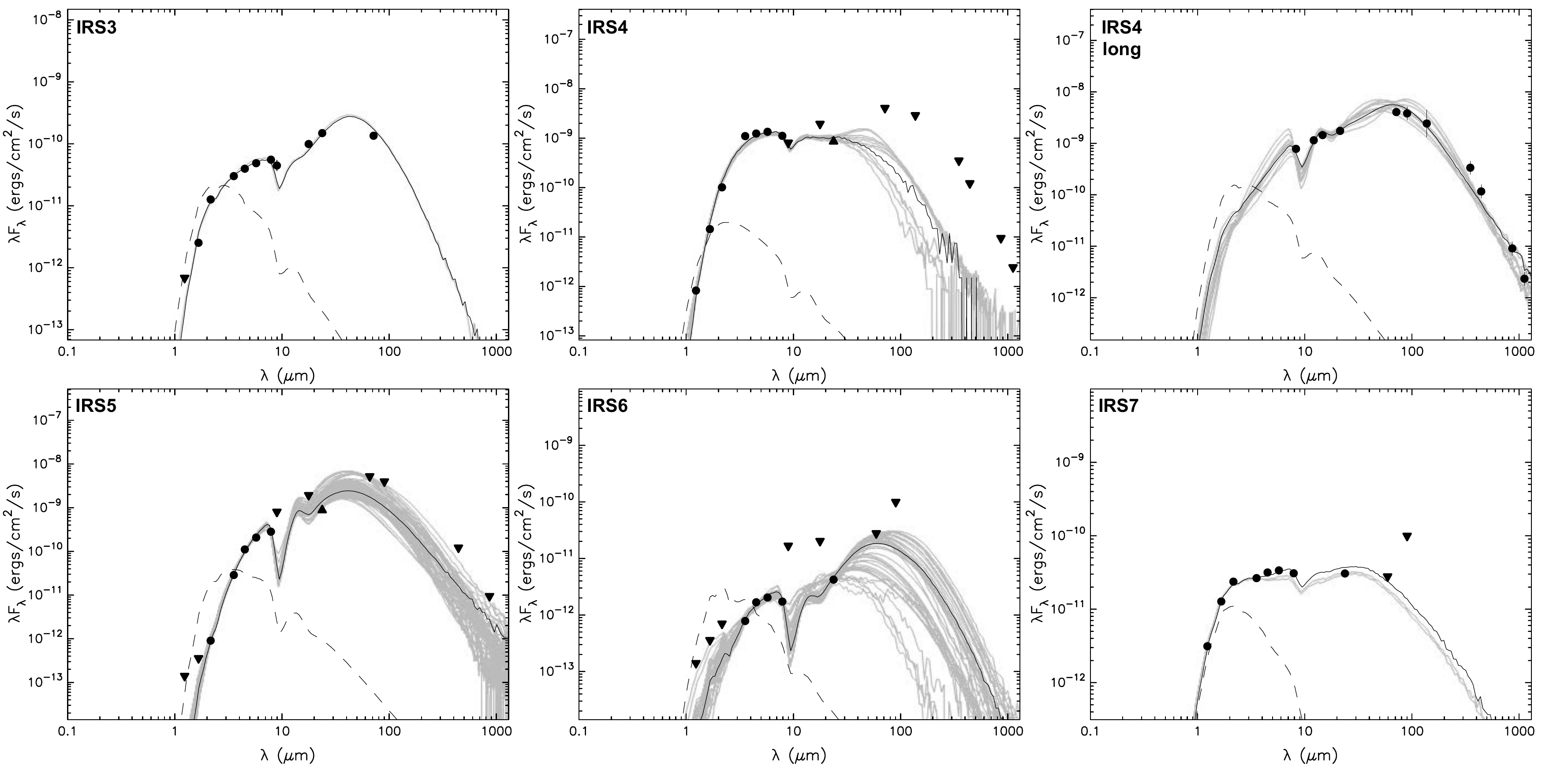}
	\caption{Spectral energy distribution (SED) of IRS3 to IRS7. Filled circles are the data points and the triangles are lower and upper limits where errors are generally smaller than the data points. The solid black line indicates the best-fitting model, and the grey lines show all models that also fit the data satisfying the $\chi^2$ - ${\chi}^2 _{best}$ $<$ 3 per data point criteria. The dashed line shows the SED of the assumed stellar photosphere in the best-fitting model.}
	\label{fig:sed-fits}
\end{figure*}

In order to better characterise the IRS objects we have explored the archive of 2D axisymmetric radiative transfer models of protostars \citep{2006ApJS..167..256R} calculated for a large range of stellar parameters (masses, accretion rates, disk masses and disk orientations), evolutionary stages (from the early envelope infall to the late disk-only) and each calculated for 10 different viewing angles developed by \citet{2007ApJS..169..328R}. The archive also provides an SED fitting interface\footnote{\url{http://caravan.astro.wisc.edu/protostars/}} which can be used to select best fit models to a given observed SED within the specified $\chi^2$. We have used this interface to input all the available magnitude/flux values for the IRS objects from various archives and databases (presented in Table\,\ref{irsphotom}). Those values were imputed assuming a conservative error of at least 10\% in order to avoid \emph{over-interpretation} of the SEDs \citep{2007ApJS..169..328R}, although the true errors are smaller in the near to mid-infrared range. The observed SED can also be scaled to match the models in the grid by using a range of values in extinction (A$_V$) and distance. For the distance we took range from 1 to 1.7 kpc but for IRS4 and 5 we narrowed down the range to 1-1.3\,kpc, since the distance to the \textbf{IR cluster} (1.17\,kpc) is known with higher certainty. For the A$_V$ we used values from  \citet{2009ApJS..184...18G} with $\pm$ 10\% range.

\begin{table*}
	\centering
	\begin{minipage}{160mm}
	\caption{Model best fit results (\textbf{in bold face text}) satisfying the $\chi^2$ - ${\chi}^2 _{best}$ $<$ 3 per data point criteria for IRS objects using online SED fitting interface by \citet{2007ApJS..169..328R}. Provided errors are calculated as difference between lowest and highest parameter values of SED models compared to the best fit satisfying the same criteria. See Section\,\ref{sec:sedmodels} for explanation and discussions.}\label{irsmodel}
	\begin{tabular}{lrrrrrr}
\hline
Parameter & IRS3 & IRS4 & IRS4-long & IRS5 & IRS6 & IRS7\\
\hline
\smallskip
$M_*$ (M$_{\sun}$)		& \textbf{1.88}$\pm^{0.00} _{0.00}$
						& \textbf{9.90}$\pm^{2.10} _{4.64}$
						& \textbf{4.49}$\pm^{1.97} _{2.57}$
						& \textbf{3.66}$\pm^{4.78} _{3.16}$
						& \textbf{0.29}$\pm^{3.43} _{0.19}$
						& \textbf{2.24}$\pm^{0.00} _{0.00}$\\
\smallskip
Age $\times 10^4$ (yr)	& \textbf{4.83}$\pm^{0.00} _{0.00}$
						& \textbf{138}$\pm^{250} _{29}$
						& \textbf{0.58}$\pm^{1.98} _{0.47}$
						& \textbf{0.22}$\pm^{70.4} _{0.10}$
						& \textbf{0.68}$\pm^{982} _{0.58}$
						& \textbf{54.6}$\pm^{0.0} _{0.0}$\\
\smallskip
$R_*$ (R$_{\sun}$)		& \textbf{10.6}$\pm^{0.0} _{0.0}$
						& \textbf{3.97}$\pm^{0.43} _{1.22}$
						& \textbf{31.1}$\pm^{9.2} _{16.3}$
						& \textbf{26.7}$\pm^{35.0} _{23.6}$
						& \textbf{4.76}$\pm^{2.54} _{3.01}$
						& \textbf{4.93}$\pm^{0.0} _{0.0}$\\
\smallskip
$T_* \times 10^3$ (K)	& \textbf{4.3}$\pm^{0.0} _{0.0}$
						& \textbf{25.5}$\pm^{2.8} _{8.5}$
						& \textbf{4.23}$\pm^{0.22} _{0.12}$
						& \textbf{4.21}$\pm^{15.09} _{0.52}$
						& \textbf{3.35}$\pm^{9.85} _{0.81}$
						& \textbf{4.96}$\pm^{0.0} _{0.0}$\\
\smallskip
$M_{disk} \times 10^{-2}$ (M$_{\sun}$) & \textbf{3.06}$\pm^{0.00} _{0.00}$
						& \textbf{7.55}$\pm^{10.85} _{7.46}$
						& \textbf{40.1}$\pm^{13.2} _{39.7}$
						& \textbf{20.40}$\pm^{19.10} _{20.39}$
						& \textbf{0.51}$\pm^{2.63} _{0.51}$
						& \textbf{10.6}$\pm^{0.0} _{0.0}$\\
\smallskip
$\dot M_{disk} \times 10^{-6}$ (M$_{\sun}$/yr) & \textbf{0.024}$\pm^{0.00} _{0.00}$
						& \textbf{0.067}$\pm^{6.333} _{0.061}$
						& \textbf{3.18}$\pm^{107.82} _{3.04}$
						& \textbf{2.88}$\pm^{94.42} _{2.87}$
						& \textbf{0.081}$\pm^{97.91} _{0.081}$
						& \textbf{1.21}$\pm^{0.0} _{0.0}$\\
\smallskip
$R_{min}$ Disk (AU)		& \textbf{2.26}$\pm^{0.00} _{0.00}$
						& \textbf{7.87}$\pm^{15.73} _{3.51}$
						& \textbf{1.13}$\pm^{11.17} _{0.24}$
						& \textbf{5.89}$\pm^{14.71} _{5.23}$
						& \textbf{0.67}$\pm^{2.32} _{0.62}$
						& \textbf{0.33}$\pm^{0.0} _{0.0}$\\
\smallskip
$R_{max}$ Disk (AU)		& \textbf{34.8}$\pm^{0.0} _{0.0}$
						& \textbf{285}$\pm^{712} _{208}$
						& \textbf{7.68}$\pm^{77.92} _{5.03}$
						& \textbf{11.9}$\pm^{553.1} _{10.39}$
						& \textbf{48.5}$\pm^{2951.5} _{46.75}$
						& \textbf{83.0}$\pm^{0.0} _{0.0}$\\
\smallskip
$M_{env.} \times 10^{-2}$ (M$_{\sun}$) & \textbf{6.57}$\pm^{0.00} _{0.00}$
						& \textbf{0.0000285}$\pm^{8.909972} _{0.000027}$
						& \textbf{572}$\pm^{13728} _{317}$
						& \textbf{100}$\pm^{2580} _{93}$
						& \textbf{35.8}$\pm^{41.5} _{35.8}$
						& \textbf{0.01}$\pm^{0.00} _{0.00}$\\
\smallskip
$\dot M_{env.} \times 10^{-6}$ (M$_{\sun}$/yr) & \textbf{8.19}$\pm^{0.00} _{0.00}$
						& \textbf{0.0}$\pm^{0.004} _{0.000}$
						& \textbf{464}$\pm^{1296} _{389}$
						& \textbf{31.9}$\pm^{101.1} _{30.9}$
						& \textbf{7.5}$\pm^{24.9} _{7.5}$
						& \textbf{0.09}$\pm^{0.058} _{0.000}$\\
\smallskip
$R_{max}$ Env. $\times 10^3$ (AU)	& \textbf{3.22}$\pm^{0.00} _{0.00}$
						& \textbf{0.0}$\pm^{20.2} _{0.0}$
						& \textbf{5.04}$\pm^{69.56} _{0.37}$
						& \textbf{8.7}$\pm^{69.7} _{6.1}$
						& \textbf{5.08}$\pm^{2.53} _{5.08}$
						& \textbf{1.13}$\pm^{0.0} _{0.0}$\\
\smallskip
$L_{tot}$ (L$_{\sun}$)	& \textbf{34.5}$\pm^{0.0} _{0.0}$
						& \textbf{6030}$\pm^{5570} _{5450}$
						& \textbf{290}$\pm^{196} _{102}$
						& \textbf{212}$\pm^{958} _{102}$
						& \textbf{2.68}$\pm^{127.32} _{2.02}$
						& \textbf{25.2}$\pm^{0.0} _{0.0}$\\
\smallskip
Incl. Angle (deg.)		& \textbf{56.6}$\pm^{6.7} _{0.0}$
						& \textbf{81.4}$\pm^{5.7} _{40.0}$
						& \textbf{18.2}$\pm^{0.0} _{0.0}$
						& \textbf{31.8}$\pm^{49.6} _{13.6}$
						& \textbf{49.5}$\pm^{37.6} _{31.3}$
						& \textbf{31.8}$\pm^{24.8} _{0.0}$\\
\smallskip
Fitted $A_V$ (mag.)		& \textbf{22.00}$\pm^{0.00} _{1.96}$
						& \textbf{22.68}$\pm^{3.32} _{0.68}$
						& \textbf{22.00}$\pm^{1.56} _{0.00}$
						& \textbf{38.76}$\pm^{11.24} _{11.69}$
						& \textbf{16.3}$\pm^{3.7} _{2.3}$
						& \textbf{14.91}$\pm^{0.00} _{0.40}$\\
\smallskip
Fitted Distance (kpc)	& \textbf{1.15}$\pm^{0.00} _{0.05}$
						& \textbf{1.12}$\pm^{0.17} _{0.12}$
						& \textbf{1.23}$\pm^{0.06} _{0.23}$
						& \textbf{1.17}$\pm^{0.11} _{0.17}$
						& \textbf{1.55}$\pm^{0.15} _{0.84}$
						& \textbf{1.17}$\pm^{0.05} _{0.00}$\\
\smallskip
N (models)				& \textbf{2}
						& \textbf{19}
						& \textbf{24}
						& \textbf{78}
						& \textbf{96}
						& \textbf{4}\\
		\hline
	\end{tabular}
\end{minipage}
\end{table*}

Figure\,\ref{fig:sed-fits} shows the SED fits for the IRS objects where the solid black line in each case represents the best fit model and the grey lines show the next best fits which satisfy the criteria of $\chi^2$ - ${\chi}^2 _{best}$ $<$ 3 per data point \citep{2007ApJS..169..328R}. In the case of IRS4 we present a second set of models which better fit the long wavelengths part of the SED (\textbf{IRS4 long}). Results from the fits are then listed in Table\,\ref{irsmodel} where first column lists the names of fitted parameters and in all subsequent columns the parameters of SED fits are given for each IRS object. 

Before discussing each object and determining how accurate the derived physical parameters are from the given SED fit, we would like to address some of the general discrepancies that one could expect from the procedure itself. We emphasise that the SED fitter is not a tool to accurately derive the physical parameters but it provides a way to determine how well constrained each of the different parameters are  \citep{2007ApJS..169..328R}. Those parameters strongly depend on the wavelength coverage for each object and are better constrained by the multi-waveband data points from the near-infrared to radio (see Table\,\ref{irsphotom} for our coverage). 

In cases when the observed data-points were \emph{enough} to fit a relatively accurate SED there have been some cases (CoKu\,Tau/1 and GG\,Tau for example) when the derived parameters where quite unrealistic when compared with the observed data and the independent theoretic analysis \citep[see][for the details]{2007ApJS..169..328R}. The SED fitter uses pre-compiled grid of model SED \citep{2006ApJS..167..256R} to derive the physical parameters and as numerous as they are it would be naive to suggest that we have a complete coverage of all the possible nuances that influence or drive the star formation process. We would like to note that it is not the scope of this work to give exhaustive explanation to all those discrepancies but rather to use these SED fits as first approximation or as \emph{a work in progress}.

\subsubsection{IRS3}
\label{sec:irs3}

IRS3 was previously reported to be a Class\,I protostar by \citet{2009ApJS..184...18G} using near to mid-infrared colour-colour analysis. Our SED modelling gives a good fit (Fig.\,\ref{fig:sed-fits}) which is largely due to the numerous multi-waveband data points (1.2 to 70 $\mu$m) present for the object. The degeneracy or number of models satisfying the best-fit criteria is just 2 with slight difference in disk inclination angle, fitted distance and the fitted A$_V$ (Tab.\,\ref{irsmodel}). The fitted distance ($\sim$1.15kpc) is in good agreement with the value inferred from the observations of the \textbf{IR cluster} \citep{1996A&A...308..573M}. The same can be said about the fitted A$_V$ ($\sim$20--22) which is also in a good agreement with the value derived by \citet{2009ApJS..184...18G}. The error in fitted inclination angle is not unexpected since this is one of the least well determined parameters in the SED fitting process but it is directly connected with the discrepancy in distance and A$_V$ determination which is the result of an input $\pm$10\% assumed range.

The rest of the fitted parameters do not show any degree of degeneracy. However, due to the fact that our data does not extend beyond the \emph{Spitzer} MIPS 70$\mu$m we suspect that some of the parameters are not well constrained. In particular the absence of any (sub)mm data points makes the precise estimates of the disk mass ($M_{disk}$) questionable and introduces a degeneracy between the central source temperature ($T_*$) and the disk accretion rate ($\dot M_{disk}$). On the other hand we can stipulate that SED fit can not be interpreted by disk-only source but there is a significant rise after $\sim$20$\mu$m which can only be interpreted by the presence of an envelope and its higher rate of accretion ($\dot M_{env.}$) \citep{2006ApJS..167..256R}. This last point strongly supports the protostellar nature of the IRS3 object confirming the findings by \citet{2009ApJS..184...18G}. The classification of this source and the positional placement in regards to MHO\,735 leaves no doubt that IRS3 is the driving source of a short (0.028pc) outflow. 

\subsubsection{IRS4 and IRS4 long}
\label{sec:irs4}

IRS4 is one of the prominent members of the \textbf{IR Cluster}. It has been observed at multiple wavelengths between 1.2 to 1300$\mu$m (as can be seen from the fitted SED (Fig.\,\ref{fig:sed-fits}). The prominence of this object enabled its early discovery as a red source called CPM\,19 which was then resolved into \textbf{IR Cluster}  \citep{1989AJ.....98..643C,1988MNRAS.232..497C,1994ApJS...94..615H}, as noted in  Section\,\ref{sec:intro}. Recently \citet{2009Ap.....52..501N} observed the same object in optical I and R bands and reported about the variability comparing DSS1, DSS2 and current data. In addition to these facts IRS4 was classified as a Class\,I protostar by \citet{2009ApJS..184...18G} based on near to mid-infrared colour-colour analysis. 

The observational facts would seem to suggest that IRS4 is quite an energetic source and it would be misleading to place it into one or other classification group. Long-wave observations suggest a possibility of high-mass source \citep{2008A&A...481..345M} which would make low-mass classification of YSOs incompatible or inaccurate. It is worth mentioning that all of the previous infrared to radio observations were done using larger apertures which would invalidate any kind of comparison with the current near to mid-infrared data due to the \textbf{IR Cluster} source contamination. With these facts in hand we performed two separate SED fittings of IRS4 object; in one SED fit we used only near to mid-infrared data obtained with a small aperture ($\sim$5000AU) and the other we used large-aperture ($\sim$22000AU) data to incorporate the flux values (such as MSX6, c.f. Tab.\,\ref{irsphotom}) from lower resolution mid-infrared to radio wavelengths (IRS4 long). Taking into consideration all the possible discrepancies in the data points it would be almost impossible to correctly derive any physical parameters. However, as a first approximation we tried to give some interpretations noting the highly speculative nature of them.

When we fit an SED to the near to mid-infrared data points (Fig.\,\ref{fig:sed-fits}) we get edge-on system containing zero age main sequence star or ZAMS of about 10M$\sun$ with almost no envelope and relatively small accreting disk. The fitted temperature would seem to indicate a spectral type of late \textbf{O} or early \textbf{B} with total luminosity of 6000 L$\sun$ which would explain its observability in all the wavelengths. In contrast, if we fit the data-points represented by larger aperture observations (Fig.\,\ref{fig:sed-fits}) we get about 4.5M$\sun$ Class\,0/I actively accreting protostar with a large envelope which would seem to fit the role of outflow-driving source. Both of these fits have considerable level of degeneracy which is understandable due to the uncertainties in the apertured used (Tab.\,\ref{irsmodel}).

This kind of clear separation between sources fitted using the different apertures suggests a possibility of multiple objects in the line of sight belonging to the same \textbf{IR Cluster}. It would seem that the visible IRS4 is situated in front of an embedded core which probably contains Class\,0/I protostar coinciding with the CM-A core (Fig.\,\ref{fig:irs4and5}). This kind of superimposition would explain the abundance of outflows from the same area but it would be difficult to stipulate which outflow belongs to which source since there is a possibility of more hidden objects along the line of sight.

\subsubsection{IRS5}
\label{sec:irs5}

IRS5 is one of the Class\,I sources identified by \citet{2009ApJS..184...18G} which is almost invisible at near-infrared wavelengths. The picture is quite different in mid to far-infrared bands where IRS5 even outshines IRS4 (24$\mu$m on Fig.\,\ref{fig:genfield}; Tab.\,\ref{irsphotom}). Our analysis of this object provides a challenge due to its apparent closeness to IRS4, preventing accurate PSF flux measurements. Similar discrepancies are also invalidating some of the flux values presented in various catalogues forcing us to use them only as upper-limits which rises the degeneracy of the SED fits (Tab.\,\ref{irsmodel}). Despite that, from the SED fit (Fig.\,\ref{fig:sed-fits}) we can derive some reasonable characterisation for IRS5.

IRS5 is one of the deeply embedded members of this \textbf{IR Cluster} of stars. Our SED fit suggests a very young Class\,0/I protostar possibly in the same stage as CM-A core discussed in Sect.\,\ref{sec:irs4}. There was no 3.6cm core reported in this position (cf. Fig.\,\ref{fig:irs4and5}) within 3$\sigma$ limit of the observations \citep{2002ApJ...570..758M} but there is a hint of a possible object below or around that limit which would need to be followed by separate observations. IRS5 could be the driving source of the MHO\,740A and B flow but those MHOs seem to be aligned with the knots in MHO\,734 and are likely being driven from the CM-B core rather than from IRS5, which is not well aligned with the axis of this MHO outflow (Fig.\,\ref{fig:irs4and5}).

\subsubsection{IRS6}
\label{sec:irs6}

One of the possible candidates of outflow driving sources in the presented region is IRS6. This object was identified as a Class\,I YSO by \citet{2009ApJS..184...18G}, based solely on its mid-IR colours (it is undetected in the near-IR). The scarceness of valid data points throughout the whole electromagnetic spectrum prevented us from producing a reliable SED fit for IRS6 as can be seen from Fig.\,\ref{fig:sed-fits} and Tab.\,\ref{irsmodel}. The best fit SED values seem to indicate a young Class\,0/I low-mass star (0.29M$\sun$) situated relatively further (1.55kpc) than the rest of the IRS objects being discussed here. All of the fitted parameters in this case are very sensitive to the range of A$_V$ used which is still uncertain for this particular source. In our fit we assumed range of 14 - 20 based on the fact that values of 14 to 20 were reported for IRS7 and IRS3 respectively by \citet{2009ApJS..184...18G} and IRS6 is situated in-between.

\subsubsection{IRS7}
\label{sec:irs7}

IRS7 is the star which illuminates the GM\,2-4 (GGD\,4b) optical reflection nebula and is the probable driving source of the MHO\,742-743 outflow (Fig.\,\ref{fig:irs7}). We have relatively well sampled data points from near to mid-infrared range of electromagnetic spectrum which enabled us to produce quite good SED fit as can be seen on Fig.\,\ref{fig:sed-fits}. Following the same logic as we did in case of IRS3 (see Sect.\,\ref{sec:irs3}) we can suggest that the best fit SED parameters (cf. Tab.\,\ref{irsmodel}) infer relatively evolved Class\,II YSO rather than Class\,I as suggested by \citet{2009ApJS..184...18G}. We find that IRS7 is a low-mass YSO with active accreting disk which is dominating the emission coming from this source. All the fitted SED parameters enforce our scenario of IRS7 being the driving source of the 742-743 bipolar outflow.

\subsection{Star Formation in the vicinity of GM\,2-4}
\label{sec:sf}

\begin{figure}
	\centering
		\includegraphics[width=8cm]{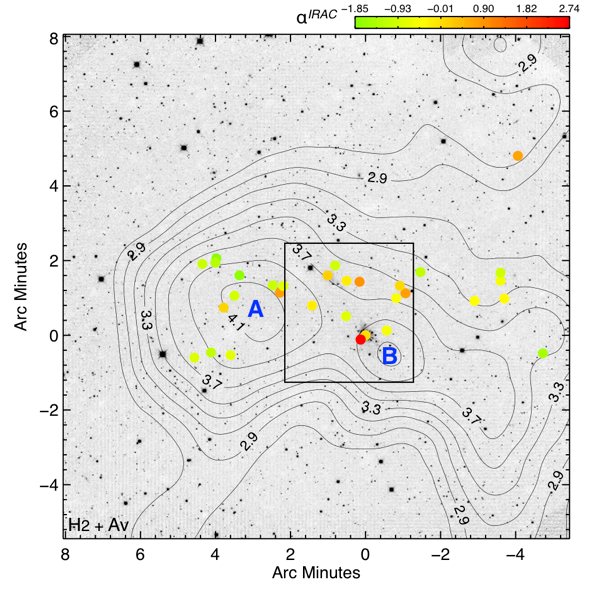}
	\caption{The wide-field view of the region in H$_2$ where offsets are measured from IRS4 and solid line box shows the area presented in Fig.\,\ref{fig:mainfigure}(b). Overlaid contours are from A$_V$ map constructed from 2MASS PSC and provided by \citet{2009MNRAS.395.1640R}. Filled circles indicate the positions of IRAC identified Class\,I/II YSO sources provided by \citet{2009ApJS..184...18G} where the colour spans according to the range of IRAC SED Slope ($\alpha^{IRAC}$).}
	\label{fig:region-final}
\end{figure}

Fig.\,\ref{fig:region-final} shows the wide-field area where the search for outflows was performed. However, MHO flows were only detected in a small sub-region presented in Fig.\,\ref{fig:mainfigure}(b). This suggests that the immediate vicinity of the IRS3 to IRS7 is the site of relatively recent star formation activity. This scenario becomes even more favourable if we consider the disk prominence of the YSOs as defined by the parameter $\alpha^{IRAC}$ (the slope of the SED across the mid-infrared wavelengths observed with the IRAC instrument on \emph{Spitzer}) and their overall distribution in East and West of the area. There are more Class\,II or weak-disk sources in the East than in the west.

Furthermore let us consider $\alpha^{IRAC}$ and the relative distribution of the YSOs to the two distinct A$_V$ peaks (\textbf{A} and \textbf{B}) shown on Fig.\,\ref{fig:region-final}. Most of the YSOs with significant circumstellar disks are associated with the A$_V$ peaks but not strictly confined to the peaks which could generally be due to the dispersal of protostars from their original birthplace showing somewhat uniform spread in relation to the A$_V$ peaks as it can be seen in case of \textbf{peak A}. This would mean that YSO population associated with \textbf{peak A} could have already gone through the active phase of star formation and therefore had time to disperse almost uniformly in relation to the \textbf{peak A} and have more evolved appearance.

The same could not be said for the YSO population associated with the \textbf{peak B} (Fig.\,\ref{fig:region-final}) since in this case there is no uniform distribution rather than the protostars are situated in the North-East part of the peak suggesting that the star formation progresses from North-East to West, South-West for the entire region discussed here. This is in good agreement with the scenario of star formation around A$_V$ \textbf{peak A} and non-detection of outflows in that area.

\section{Summary}
\label{sec:sum}

We have presented near-infrared H$_2$ 1-0 S(1) line study of the region surrounding GM\,2-4 alternatively known as GGD\,4 in order to reveal the outflow content of the area and to better understand the earlier suspected star formation processes in this area. In our quest to understand this region we combined multi-waveband data-sets obtained from various online archives. Our findings and analysis can be summarised as following:
\begin{itemize}
	\item The analysis of our near-infrared H$_2$ 1-0 S(1) line data enabled us to identify 27 individual shocked features which combine into 5 bipolar and 2 monopolar outflows driven from the protostars in the immediate vicinity of the GM\,2-4 reflection nebula. We were able to cross-correlate some of the H$_2$ identified objects with previously reported optical HH objects as well as with \emph{green fuzzy} features in \emph{Spitzer} IRAC 4.5$\mu$m wide-field band thought to have a shocked nature.
	\item We were able to propose the driving sources for the detected outflows based on the morphological and positional characteristics of the individual shocked features. All of the driving sources were identified as YSO based on the previously available knowledge \citep{2009ApJS..184...18G,2009Ap.....52..501N} and our SED modelling. In one particular case, IRS4, our SED modelling enabled us to suggest a solution for the previously known discrepancy \citep{1996A&A...308..573M,2002ApJ...570..758M} as two different stars one late \textbf{O} to early \textbf{B} type 10M$\sun$ ZAMS high mass star and the other as very early, possibly Class\,0 protostar on the same line of sight which might be the actual driving source of the several bipolar outflows.
	\item Our analysis of the wide area using IRAC SED Slope in conjunction with the A$_V$ map including the GM\,2-4 suggests that star formation progresses roughly from North-East to West, South-West. This result confirms the conclusion made in our previous paper \citep{2009Ap.....52..501N} mainly on the base of the optical data.
\end{itemize}

There are still several facts that we wish to follow-up in order to understand the exact outflow dynamics involved in the \textbf{IR cluster} for which further high-resolution imaging, spectroscopy and radio-line data would be beneficiary. The importance of spectroscopic followup and monitoring of  IRS4 (CPM19) would prove useful in determining the exact nature of variability in this object.

\section*{Acknowledgments}

TK acknowledges support of the Science Foundation Ireland (SFI) Research Frontiers award. This work was partly supported by INTAS grant 03-51-4838 and ANSEF grant PS-astroex 2517. This publication makes use of data products from the Two Micron All Sky Survey, which is a joint project of the University of Massachusetts and the Infrared Processing and Analysis Center/California Institute of Technology, funded by the National Aeronautics and Space Administration and the National Science Foundation. This research has also made use of the NASA/ IPAC Infrared Science Archive, which is operated by the Jet Propulsion Laboratory, California Institute of Technology, under contract with the National Aeronautics and Space Administration.

\bibliographystyle{mn2e}
\bibliography{i05373p2349}
\appendix

\section{Photometry}

\begin{table*}
	\caption{Photometric value database for all IRS object converted into mJy with the appropriate errors and used apertures to model SEDs. Origin of the each data-set is indicated in the last column.}
	\label{irsphotom}
	\begin{tabular}{lrrrrrr}
        \hline
Filter 
		& IRS3$\pm$Err. (Ap.$\arcsec$)
		& IRS4$\pm$Err. (Ap.$\arcsec$)
		& IRS5$\pm$Err. (Ap.$\arcsec$)
		& IRS6$\pm$Err. (Ap.$\arcsec$)
		& IRS7$\pm$Err. (Ap.$\arcsec$)
		& Data$^{\dagger}$\\ 
\hline
2MASS J 
		& $<$\textbf{0.282} (3)
		& \textbf{0.344}$\pm$0.046 (5)
		& $<$\textbf{0.058} (4)
		& $\ldots$
		& \textbf{1.294}$\pm$0.059 (4)
		& (1)\\
2MASS H
		& \textbf{1.400}$\pm$0.087 (3)
		& \textbf{8.029}$\pm$0.247 (5)
		& $<$\textbf{0.199} (4)
		& $\ldots$
		& \textbf{7.065}$\pm$0.230 (4)
		& (1)\\
2MASS K
		& \textbf{9.118}$\pm$0.329 (3)
		& \textbf{73.236}$\pm$2.127 (5)
		& \textbf{0.661}$\pm$0.086 (4)
		& $\ldots$
		& \textbf{17.074}$\pm$0.465 (4)
		& (1)\\
IRAC Ch1
		& \textbf{35.690}$\pm$0.972 (5)
		& \textbf{1307.831}$\pm$11.990 (5)
		& \textbf{34.084}$\pm$0.622 (4)
		& \textbf{0.930}$\pm$0.008 (4)
		& \textbf{31.373}$\pm$0.287 (5)
		& (2),(3)\\
IRAC Ch2
		& \textbf{59.504}$\pm$2.152 (5)
		& \textbf{1864.438}$\pm$17.093 (5)
		& \textbf{166.935}$\pm$1.530 (4)
		& \textbf{2.527}$\pm$0.023 (4)
		& \textbf{47.703}$\pm$0.437 (5)
		& (2),(3)\\
IRAC Ch3
		& \textbf{93.046}$\pm$3.365 (5)
		& \textbf{2586.412}$\pm$23.712 (5)
		& \textbf{398.747}$\pm$3.656 (4)
		& \textbf{3.915}$\pm$0.036 (4)
		& \textbf{64.967}$\pm$0.596 (5)
		& (2),(3)\\
IRAC Ch4
		& \textbf{145.566}$\pm$3.967 (5)
		& \textbf{2931.306}$\pm$26.874 (5)
		& \textbf{743.124}$\pm$6.813 (4)
		& \textbf{4.561}$\pm$0.124 (4)
		& \textbf{81.482}$\pm$0.747 (5)
		& (2),(3)\\
MSX6 A
		& \textbf{193.8}$\pm${13.6} (15)
		& $<$\textbf{2169.7} (15)
		& $\ldots$
		& $\ldots$
		& $\ldots$
		& (4)\\
AKARI/IRC 9
		& \textbf{135.82}$\pm$0.223 (5)
		& \textbf{2411.56}$\pm$175.4 (8)
		& $\ldots$
		& $\ldots$
		& $\ldots$
		& (5)\\
IRAS 12
		& $\ldots$
		& $<$\textbf{9020} (30)
		& $\ldots$
		& $\ldots$
		& $\ldots$
		& (6)\\
MSX6 C
		& $\ldots$
		& $<$\textbf{4652.9} (15)
		& $\ldots$
		& $\ldots$
		& $\ldots$
		& (4)\\
MSX6 D
		& \textbf{798.1}$\pm$79.8 (15)
		& $<$\textbf{7098.3} (15)
		& $\ldots$
		& $\ldots$
		& $\ldots$
		& (4)\\
AKARI/IRC 18
		& \textbf{588.53}$\pm$57.9 (6)
		& \textbf{11499.8}$\pm$53.4 (10)
		& $\ldots$
		& $\ldots$
		& $\ldots$
		& (5)\\
MSX6 E
		& $\ldots$
		& $<$\textbf{12474} (15)
		& $\ldots$
		& $\ldots$
		& $\ldots$
		& (4)\\
MIPS 24
		& \textbf{1174.08}$\pm$31.99 (7)
		& $>$\textbf{6768.59} (4)
		& $>$\textbf{6945.42} (4)
		& \textbf{33.55}$\pm$1.21 (5)
		& \textbf{243.05}$\pm$2.23 (7)
		& (2),(3)\\
IRAS 25
		& $\ldots$
		& $<$\textbf{305560} (30)
		& $\ldots$
		& $\ldots$
		& $\ldots$
		& (6)\\
IRAS 60
		& $\ldots$
		& $<$\textbf{144830} (30)
		& $\ldots$
		& $\ldots$
		& $\ldots$
		& (6)\\
AKARI/FIS 65
		& $\ldots$
		& \textbf{115140}$\pm$7332 (30)
		& $\ldots$
		& $\ldots$
		& $\ldots$
		& (7)\\
MIPS 70
		& $<$\textbf{3223} (30)
		& $<$\textbf{96926} (30)
		& $\ldots$
		& $\ldots$
		& $\ldots$
		& (3)\\
AKARI/FIS 90
		& $\ldots$
		& $<$\textbf{120067} (30)
		& $\ldots$
		& $\ldots$
		& $\ldots$
		& (7)\\
IRAS 100
		& $\ldots$
		& $<$\textbf{191160} (38)
		& $\ldots$
		& $\ldots$
		& $\ldots$
		& (6)\\
AKARI/FIS 140
		& $\ldots$
		& \textbf{132767}$\pm$23968 (30)
		& $\ldots$
		& $\ldots$
		& $\ldots$
		& (7)\\
AKARI/FIS 160
		& $\ldots$
		& \textbf{239313}$\pm$3880.5 (30)
		& $\ldots$
		& $\ldots$
		& $\ldots$
		& (7)\\
350$\mu$m
		& $\ldots$
		& $<$\textbf{41000} (30)
		& $\ldots$
		& $\ldots$
		& $\ldots$
		& (8)\\
SCUBA 450$\mu$m
		& $\ldots$
		& $<$\textbf{18000} (30)
		& $\ldots$
		& $\ldots$
		& $\ldots$
		& (8)\\
SCUBA 850$\mu$m
		& $\ldots$
		& $<$\textbf{2730} (35)
		& $\ldots$
		& $\ldots$
		& $\ldots$
		& (8)\\
1.1mm
		& $\ldots$
		& $<$\textbf{910} (30)
		& $\ldots$
		& $\ldots$
		& $\ldots$
		& (8)\\
1.3mm
		& $\ldots$
		& $<$\textbf{700} (30)
		& $\ldots$
		& $\ldots$
		& $\ldots$
		& (8)\\
		\hline
\end{tabular}
	\par $^\dagger$ Data origin: (1) 2MASS by \citet{2006AJ....131.1163S}; (2) \citet{2009ApJS..184...18G}; (3) this publication; (4) MSX6 PSC by \citet{2003yCat.5114....0E}; (5) AKARI/IRC by \citet{2010A&A...514A...1I}; (6) IRAS PSC; (7) AKARI/FIS by \citet{2007PASJ...59S.389K}; (8) \citet{2000A&A...355..617M}.
\end{table*}

\bsp
\label{lastpage}
\end{document}